# Femtosecond laser Fourier transform absorption spectroscopy

Julien Mandon, Guy Guelachvili, Nathalie Picqué
Laboratoire de Photophysique Moléculaire, CNRS; Univ. Paris-Sud, Bâtiment 350, 91405 Orsay, France

Frédéric Druon, Patrick Georges
Laboratoire Charles Fabry, Institut d'Optique Graduate School, CNRS; Campus Polytechnique, RD 128, 91127 Palaiseau Cedex

Corresponding author:
Dr. Nathalie Picqué,
Laboratoire de Photophysique Moléculaire
Unité Propre du CNRS, Université de Paris-Sud, Bâtiment 350
91405 Orsay Cedex, France
Phone nb: 33 1 69 15 66 49
Fax nb: 33 1 69 15 75 30
Email: nathalie.picque@ppm.u-psud.fr
Web: http://www.laser-fts.org

Abstract: A femtosecond mode-locked laser is used for the first time as a broadband infrared source for high resolution Fourier transform absorption spectroscopy. Demontration is made with a $Cr^{4+}$:YAG laser. The entire $\nu_1+\nu_3$ vibration-rotation band region of acetylene, observed after passing through a single pass 80-cm long cell, is simultaneously recorded between 1480 and 1600 nm, in 7.9 s with a signal to noise ratio equal to 1000. Two hot bands of the most abundant acetylene isotopologue and the $\nu_1+\nu_3$ band of the $^{13}C^{12}CH_2$ are also present. The noise equivalent absorption coefficient at one second averaging is equal to $7\ 10^{-7}$ $cm^{-1}.Hz^{-1/2}$ per spectral element.

OCIS codes: 300.6390 Spectroscopy, molecular, 300.6300 Spectroscopy, Fourier transforms, 140.4050 Mode-locked lasers, 140.3070 Infrared and far-infrared lasers, 140.7090 Ultrafast lasers

Over the past few years, high sensitivity spectroscopic techniques have gained in efficiency and several experimental schemes now enable to reach ppb detection levels. One of the prevailing quests presently aims at combining these sensitivities with rapid diagnostic and/or multi-species gas analysis. This induces new approaches able to couple broadband coherent sources to multiplex or multichannel spectrometers. Nowadays at high spectral resolution, simultaneous spectral coverage is still often restricted [1-3] to a few nanometers due to multichannel grating spectrometer limitation and/or to narrow spectral range optical source capabilities. With the breakthrough of mode-locked femtosecond (fs) lasers and frequency combs [4], extremely broad "steady state" high quality coherent sources have become available. This prompted the need for related spectral approaches able to deliver at once high resolution, high accuracy, broad spectral coverage and rapid acquisition. Purposely, a broadband frequency comb was coherently coupled [5] to a high finesse cavity across 100 nm, with an equivalent absorption path equal to 1.13 km. Nevertheless, due to lack of appropriate spectroscopic method, this recent impressive experiment provided simultaneous spectral results limited to 15 nm, 25 GHz resolution with signal to noise ratio (SNR) of the order of 15. Finally with the same purpose, a broadband cw $Cr^{2+}$:ZnSe laser used for intracavity absorption spectroscopy was coupled [6] to a high resolution time-resolved Fourier transform interferometer. A simultaneous coverage of 125 nm was obtained with 330 MHz resolution, 2.5 km equivalent absorption path length and SNR equal to 100.

With the above motivations in mind, a mode-locked femtosecond laser source has been efficiently analyzed by a high resolution commercial Fourier transform spectrometer. This paper reports this first-step experiment with the illustrating $C_2H_2$ spectra covering 118 nm in the 1540 nm region. The perspectives in terms of high sensitivity spectroscopy, especially in the infrared region, are then examined.

The present experiment is based on the coupling of a $Cr^{4+}$:YAG mode-locked laser source emitting in the 1.45 µm region to a Fourier transform spectrometer. $Cr^{4+}$:YAG lasers have the ability to produce pulses as short as 20 fs [7] in the wavelength range from 1250 to 1600 nm. These femtosecond lasers operate at room temperature, may be diode-pumped [8] and have larger gain bandwidths than Er-doped fiber lasers. The experimental set-up is schematized on Fig. 1. A 1064 nm commercial Nd:$YVO_4$ laser pumps a 20-mm-long $Cr^{4+}$:YAG Brewster-cut crystal. Its beam is focused onto the crystal through the dichroic mirror M1 by a 75-mm-focal-length lens. The crystal is temperature-stabilized with a thermo-electric cooler to about 286 K. The astigmatically-compensated X-shaped cavity consists of two spherical mirrors M1 and M2 with 100 mm radius of curvature (ROC), four plane chirped mirrors (CM) from Layertec M3, M4, M6, M7, a 0.8 %-transmission plane output coupler (OC), a concave focusing mirror M5 with ROC 100 mm and a semiconductor saturable absorber mirror (SESAM) from Batop on a heat sink. The CM have second-order dispersion of approximately -100 $fs^2$ and third-order dispersion of -800 $fs^3$ per bounce at 1500 nm. They provide high reflectivity (> 99.9%) in the range 1300-1730 nm. The two collimated arms respectively are 45 and 41 cm long. The cavity is in the open air. Laser threshold occurs at about 2.0 W of pumping power and 5.5 W (about 60 % absorbed) are enough to get stable mode-locked operation with pulse repetition rate of about 150 MHz and an average power of 150 mW typically measured at the output of the oscillator.

The laser beam passes through a 80-cm cell filled with acetylene in natural abundance (pressure equal to 128 hPa). It is analyzed by a commercial rapid-scan Fourier transform spectrometer (Bruker IFS66) as if it was a steady-state source. The spectrometer is equipped with a fluorine beam-splitter and an InGaAs detector. High resolution may be reached using a femtosecond laser source since the narrow molecular absorbing lines enhance the light coherence length up to a resolution limited by their linewidth. Figure 2 gives an illustration of the resulting spectra. The whole spectral domain of laser emission extends over 118 nm (500 $cm^{-1}$) with a full width at half maximum of 46 nm (197 $cm^{-1}$). Since the pulse duration is not crucial in our experiment, no autocorrelation measurement of the pulse width was performed.

Nevertheless, assuming the classical Fourier transform-limited sech shape, we estimate a pulse duration of 54 fs for a spectral bandwidth of 46 nm at 1530 nm. The spectral bandwidth restriction on the lower wavelength side comes from the SESAM reflectivity. Resolution is limited by the spectrometer to 6 GHz (0.2 cm$^{-1}$). The number of spectral elements is consequently equal to 2500. Total recording time is 7.9 s: ten scans are co-added and the fringes are scanned at a 10 kHz rate. SNR, estimated from the rms noise, is of the order of 1000. Noise equivalent absorption coefficient at one second averaging is 7 10$^{-7}$ cm$^{-1}$.Hz$^{-1/2}$ per spectral element.

All absorption lines shown on Fig.2 are due to acetylene. The most intense band is the $\nu_1+\nu_3$ band of $^{12}C_2H_2$. This acetylene isotopologue belongs to the point group $D_{\infty h}$. Since its pair of identical nuclei with a nuclear spin different from zero ( $I(H) = ½$ ), follows the Fermi statistics, even and odd rotational levels have statistical weights in the ratio of 1/3. Thanks to the high resolution revealing the rotational structure, the intensity alternation for the main band is obvious. A small part of the spectrum with an extended frequency scale is given on Fig. 3. with spectral assignments. Due to the good SNR, the $\nu_1+\nu_3$ band of H$^{12}$C$^{13}$CH is clearly seen even with a relative concentration decreased by about 100. The absence of intensity alternation due to the break of the molecular symmetry of this isotopologue is also revealed on the figure.

The coupling of Fourier transform spectrometers to mode-locked sources offers new perspectives for high resolution and high sensitivity broadband absorption spectroscopy. This experiment may indeed be improved in terms of sensitivity, spectral domain, resolution and acquisition time, as discussed below.

Sensitivity may be enhanced by increasing the interaction path between light and sample. The most promising spectra should result from the coupling demonstrated in [5] of a femtosecond frequency comb to a high finesse cavity. More precisely using a 39.4 cm long cell, corresponding to 380 MHz free spectral range, [5] reports an integrated absorption equal to 2.5 10$^{-5}$ with 1.4 ms acquisition time. Their grating spectrometer has 15 nm bandwidth (from 809 to 824 nm) at 25GHz resolution, corresponding to 270 spectral elements. Therefore their noise equivalent absorption coefficient at one second averaging is 1.4 10$^{-9}$ cm$^{-1}$.Hz$^{-1/2}$ per spectral element. Their set-up involves a cavity with a finesse equal to 4500, which corresponds to about 1.13 km equivalent path length. Absorption path of 80 m is rather common in the laboratory with a classical multipass cell. It is therefore worth noticing that a 100-time increase of the present absorption path length (80 cm) would already provide an integrated absorption equal to 10$^{-5}$ and a noise equivalent absorption coefficient at one second averaging of 7 10$^{-9}$ cm$^{-1}$.Hz$^{-1/2}$ per spectral element. With a much simpler experimental set-up, this integrated absorption would be 2.5 times better than the one reported in [5] with only 5 times worse noise equivalent absorption at one second averaging per spectral element.

Any spectral domain, from the UV to the far-infrared, may be probed with Fourier transform spectroscopy, most often with Doppler-limited resolution and with no other restriction on the spectral simultaneous coverage than the detector responsivity. Typically, up to several 10$^6$ independent spectral elements are sampled within one experiment. The technique is therefore appropriate even for supercontinua laser sources [9].

Resolving ability of the grating instruments suffer from drastic limitations when compared to Fourier interferometers. This is due to the restricted size of the available gratings and the inability of a laser beam to cover the entire grating so to benefit from its full resolution. Commercially available Fourier spectrometers based on Michelson interferometers exhibit up to 30 MHz resolution. When using a frequency comb [10] as a Fourier spectrometer, this resolution has virtually no practical limitation. For instance, 1 MHz-repetition rate mode-locked oscillators have been demonstrated [11] or stroboscopic approaches may be implemented.

Very short acquisition times are reachable with spectrometers presenting no moving parts like for instance grating, virtually imaged phased array [12], stationary Fourier interferometers [13] all equipped with CCD array detector, and frequency comb Fourier spectrometer. The frequency comb approach has the advantage of needing a single detector and therefore, in principle, does not suffer as the other techniques from a severe restriction on the number of spectral elements. This limitation may be overcome at the reasonable expenses of acquisition time. In the ms acquisition time range, ultrafast scanning Fourier Michelson interferometers are available [14]. If high resolution combined with high spectral bandwidth is needed, conventional fast-scanning Michelson interferometers appear as being the best solution: optical velocities in path difference are generally between 0.1 and 10 cm.s$^{-1}$, so that 75 MHz resolution may be reached with $10^6$ spectral elements within only 20 seconds. Time resolution ability with high spectral resolution is also at work [15].

In conclusion, we report the coupling for the first time of a femtosecond mode-locked laser used as a broadband infrared source to a high resolution Fourier transform spectrometer. Acetylene absorption spectra, recorded using a Cr$^{4+}$:YAG laser, clearly show the efficiency of this approach. Comparisons to recent experiments motivated by similar objectives are satisfactory. Improvements in sensitivity, spectral extension, resolution and acquisition time by combining femtosecond broadband laser sources to Fourier spectrometers are in progress.

Useful advices from Drs E. Sorokin and I.T. Sorokina (TU Wien, Austria) are warmly acknowledged. We thank Dr. A.V. Shestakov (E.L.S. Polyus) for supplying the Cr$^{4+}$:YAG crystal. We are grateful to L. Berger and A. Szwec for technical assistance. This work is accomplished in the frame of the Programme Pluri-Formation de l'Université de Paris-Sud "Détection de traces de gaz" 2006-2009.

Figure captions

Fig. 1.
Schematic of the experiment. The Cr[4+]YAG femtosecond laser probes a cell filled with acetylene and the resulting radiation is analysed by a Fourier transform spectrometer (FTS).

Fig. 2.
Overtone spectrum of the acetylene molecule in the 1.5 µm region demonstrating the spectral bandwidth and signal-to-noise ratio capabilities of this spectrometric technique. The strongest spectral features are the *P* and *R* branches of the $\nu_1+\nu_3$ band of $^{12}C_2H_2$.

Fig. 3
Restricted portion of the spectrum shown on Fig. 2 exhibiting rotational lines of the P branches of the $\nu_1+\nu_3$ cold band, $\nu_1+\nu_3+\nu_4^1-\nu_4^1$ and $\nu_1+\nu_3+\nu_5^1-\nu_5^1$ hot bands of $^{12}C_2H_2$ [16] and the $\nu_1+\nu_3$ cold band of $^{12}C^{13}CH_2$.

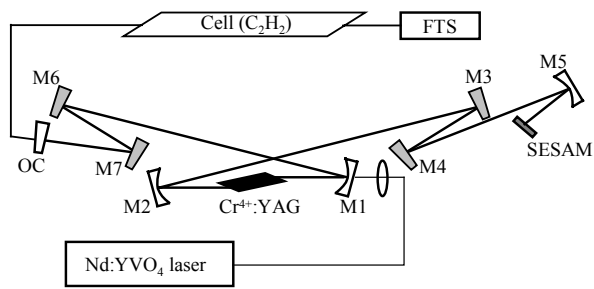

Fig. 1. Schematic of the experiment. The Cr$^{4+}$YAG femtosecond laser probes a cell filled with acetylene and the resulting radiation is analysed by a Fourier transform spectrometer (FTS).

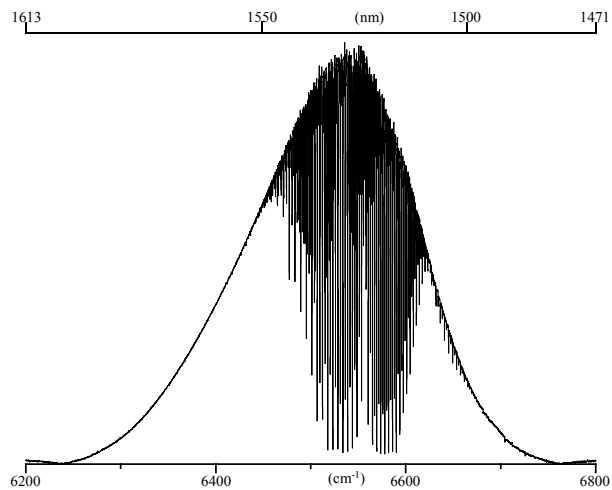

Fig. 2: Overtone spectrum of the acetylene molecule in the 1.5 μm region demonstrating the spectral bandwidth and signal-to-noise ratio capabilities of this spectrometric technique. The strongest spectral features are the $P$ and $R$ branches of the $\nu_1+\nu_3$ band of $^{12}C_2H_2$.

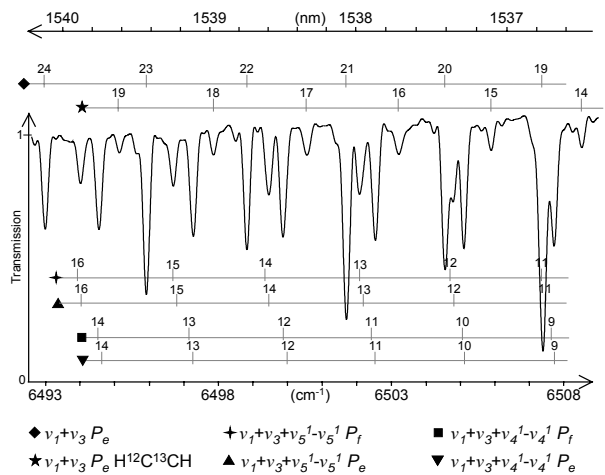

Fig. 3: Restricted portion of the spectrum shown on Fig. 2 exhibiting rotational lines of the P branches of the $\nu_1+\nu_3$ cold band, $\nu_1+\nu_3+\nu_4^1-\nu_4^1$ and $\nu_1+\nu_3+\nu_5^1-\nu_5^1$ hot bands of $^{12}C_2H_2$ [16] and the $\nu_1+\nu_3$ cold band of $^{12}C^{13}CH_2$.

References

[1] E. R. Crosson, P. Haar, G. A. Marcus, H. A. Schwettman, B. A. Paldus, T. G. Spence, and R. N. Zare, Pulse-stacked cavity ring-down spectroscopy, Review of Scientific Instruments 70, 4-10 (1999).

[2] S.M. Ball, R.L. Jones, Broadband cavity ring-down spectroscopy, Chemical Reviews 103, 5239-5262 (2003).

[3] T. Gherman, S. Kassi, A. Campargue, D. Romanini, Overtone spectroscopy in the blue region by cavity-enhanced absorption spectroscopy with a mode-locked femtosecond laser : application to acetylene, Chemical Physics Letters 383, 353-358 (2004).

[4] T. Udem, R. Holzwarth, T.W. Hänsch, Optical frequency metrology, Nature 416, 233-237 (2002).

[5] M.J. Thorpe, K.D. Moll, R.J. Jones, B. Safdi and J. Ye, Broadband Cavity Ringdown Spectroscopy for Sensitive and Rapid Molecular Detection, Science, 311, 1595-1599 (2006).

[6] N. Picqué, F. Gueye, G. Guelachvili, E. Sorokin, I.T. Sorokina, Time-resolved Fourier transform intracavity spectroscopy with a $Cr^{2+}$:ZnSe laser, Optics Letters **30**, 3410-3412 (2005).

[7] D. J. Ripin, C. Chudoba, J. T. Gopinath, J. G. Fujimoto, E. P. Ippen, U. Morgner, F. X. Kärtner, V. Scheuer, G. Angelow, T. Tschudi, Generation of 20-fs pulses by a prismless $Cr^{4+}$:YAG laser, Optics Letters 27, 61-63 (2002).

[8] S. Naumov, E. Sorokin, I. T. Sorokina, Directly diode-pumped Kerr-lens mode-locked $Cr^{4+}$:YAG laser, Optics Letters 29, 1276-1278 (2004).

[9] J.M. Dudley, G. Genty, S. Coen, Supercontinuum generation in photonic crystal fiber, Reviews of Modern Physics 78, 1135-1184 (2006).

[10] F. Keilmann, C. Gohle, R. Holzwarth, Time-domain mid-infrared frequency-comb spectrometer, Optics Letters 29, 1542-1544 (2004).

[11] D. N. Papadopoulos, S. Forget, M. Delaigue, F. Druon, F. Balembois, and P. Georges, "Passively mode-locked diode-pumped Nd:YVO$_4$ oscillator operating at an ultralow repetition rate ," Opt. Lett. 28, 1838-1840 (2003)

[12] S. Xiao and A.M. Weiner, 2-D wavelength demultiplexer with potential for ≥ 1000 channels in the C-band, Optics Express 12, 2895-2902 (2004)

[13] P. Voge, J. Primot, Simple infrared Fourier transform spectrometer adapted to low light level and high-speed operation, Optical Engineering 37, 2459-2466 (1998)

[14] P.R. Griffiths, B.L. Hirsche, C.J. Manning, Ultra-rapid-scanning Fourier transform infrared spectrometry, Vibrational Spectroscopy 19, 165–176 (1999).

[15] N. Picqué and G. Guelachvili, High-information time-resolved Fourier transform spectroscopy at work, Applied Optics 39, 3984-3990 (2000).

[16] Q. Kou, G. Guelachvili, M. Abbouti Temsamani, and M. Herman, The absorption spectrum of $C_2H_2$ around $\nu_1+\nu_3$: energy standards in the 1.5 μm region and vibrational clustering, Can.J.Phys.**72,** 1241-1250 (1994).